\documentclass[conference]{IEEEtran}
\IEEEoverridecommandlockouts
\usepackage{cite}
\usepackage{amsmath,amssymb,amsfonts}
\usepackage{algorithmic}
\usepackage{graphicx}
\usepackage{textcomp}
\usepackage{xcolor}
\usepackage{url}
\def\BibTeX{{\rm B\kern-.05em{\sc i\kern-.025em b}\kern-.08em
    T\kern-.1667em\lower.7ex\hbox{E}\kern-.125emX}}
\begin{document}

\title{
On Medical Device Software CE Compliance and Conformity Assessment
}

\author{\IEEEauthorblockN{Tuomas Granlund}
\IEEEauthorblockA{\textit{Solita Oy}\\
Tampere, Finland \\
tuomas.granlund@solita.fi}
\and
\IEEEauthorblockN{Tommi Mikkonen}
\IEEEauthorblockA{\textit{University of Helsinki}\\
Helsinki, Finland \\
tommi.mikkonen@cs.helsinki.fi}
\and
\IEEEauthorblockN{Vlad Stirbu}
\IEEEauthorblockA{\textit{CompliancePal}\\
Tampere, Finland\\
vlad.stirbu@compliancepal.eu}
}

\maketitle

\begin{abstract}
Manufacturing of medical devices is strictly controlled by authorities, and manufacturers must conform to the regulatory requirements of the region in which a medical device is being marketed for use. In general, these requirements make no difference between the physical device, embedded software running inside a physical device. or software that constitutes the device in itself. As a result, standalone software with intended medical use is considered to be a medical device. Consequently, its development must meet the same requirements as the physical medical device manufacturing. This practice creates a unique challenge for organizations developing medical software. In this paper, we pinpoint a number of regulatory requirement mismatches between physical medical devices and standalone medical device software. The view is based on experiences from industry, from the development of all-software medical devices as well as from defining the manufacturing process so that it meets the regulatory requirements.
\end{abstract}

\section{Introduction}\label{sec:introduction}
The safety of health care, medical devices, and people are among the main concerns of governments. Therefore, manufacturing of medical devices is strictly controlled by authorities, and manufacturers must conform to the regulatory requirements of the region in which a medical device is being marketed for use. Within the EU region, European Commission runs a regulatory framework, and medical devices are currently regulated by three directives, which will be replaced by two new regulations. The new regulations will fully apply after a transitional period. Based on the current time limits, the transitional period lasts until May 2020 for medical devices and until May 2022 for in vitro diagnostic medical devices. 

In the context of medical device software (MDSW), a large number of products are currently being regulated by the Council Directive 93/42/EEC on Medical Devices (MDD)\cite{mdd} and its latest amendment \cite{mdd_amd}. After the transitional period they will be regulated by Regulation (EU) 2017/745 on medical devices (MDR) \cite{mdr}. Other directives and regulations may also apply, depending on the type of software product.

Neither MDD nor MDR differentiates between the physical device, embedded software running inside a physical device or software that is the device itself \cite{mdd,mdr}. As a result, standalone software with intended medical use is considered to be a medical device, and its development must follow the same directives, regulations, national laws, and technical standards as the physical medical device manufacturing. This practice creates a unique challenge for organizations developing medical software as it is evident that regulatory requirements were created with physical devices in mind.

In this paper, we point out a number of regulatory requirement mismatches between physical medical devices and standalone medical device software. The challenges related to these mismatches will become even more critical in the near future when MDR is fully applied, and a large number of existing medical software products will be up-classified to higher classes. Consequently, as there will be hardly any software-only products classified below Class IIa \cite{mdcg2019-11}, the medical software manufacturers must have a Quality Management System (QMS) established and involve the Notified Body in their certification process in order to be able to place the products on the EU market.

The rest of this paper is structured as follows. In Section \ref{sec:background}, we provide the background for the paper. In Section \ref{sec:mismatch}, which forms the core of the paper, we present some obvious mismatches between hardware-oriented regulatory requirements and software. In Section \ref{sec:discussion}, we give a brief discussion of our observations. Finally, in Section \ref{sec:conclusions}, we draw some conclusions.

\section{Background} \label{sec:background}
In this section, we present an overview of EU regulation and its different layers. Furthermore, the processes of CE marking and placing on the EU market is discussed. 

\subsection{EU regulatory framework overview}
The regulatory framework in the EU can be interpreted to consist of several layers, including the following: 
\begin{itemize}
\item Union harmonized legislation (incl. essential requirements), 
\item National legislation,
\item Harmonized standards,
\item Guidelines. 
\end{itemize}
EU harmonized legislation includes legal acts of directives and regulations. There is a significant difference between the two: regulations will enter into force directly in all member states, whereas directives define a specific set of objectives that member states must fulfill, for example, with national legislation. The latter approach leaves a certain amount of leeway to member states, and it may result in variations in national practices. However, it is also possible to supplement EU regulations with national laws. 

The EU Commission has noticed certain shortcomings in the medical device directives \cite{stahlberg2015} and, as discussed in Section \ref{sec:introduction}, there is an on-going transitional period from current directives to corresponding new regulations. Since the first entry into force on April 2017, both regulations have been corrected twice with a specific Corrigendum. It is worth to notice that the 2\textsuperscript{nd} Corrigendum for MDR included  a surprisingly significant change concerning to extended "grace period" for those devices that were classified as Class I devices by MDD and that will be up-classified by MDR. Overall, the transition process has demonstrated that changing the EU-wide legal framework is a long and complicated process. 

An essential feature of Union harmonization legislation has been limiting the legislation to product-related essential requirements that protect the health and safety of users and stakeholders \cite{blueguide2016}. In the context of medical devices, essential requirements ensure that medical devices are effective, safe to use, and are able to achieve their intended use. However, a certain amount of technical details have been left to harmonized standards.

European standards are technical specifications, adopted first by a recognized standardization body -- such as ISO (https://www.iso.org/) and IEC (https://www.iec.ch/) -- and then by the European standardization organizations. Harmonized standards are European standards adopted, by request made by the EU Commission, for application of Union harmonization legislation \cite{blueguide2016}. They usually include informative annex that indicates the covered legal requirements and possible limitations of the coverage. Harmonized standards provide a presumption of conformity with the essential requirements they aim to cover. While the application of harmonized standards is voluntary for the manufacturer, they are, in fact, an effective tool to demonstrate conformity to regulatory requirements. 

The EU Commission has mechanisms to provide a range of non-binding guideline documents in order to support uniform application of the directives and regulations within the EU. These guidance documents are usually developed in active conjunction with the industry stakeholders. In the MDD context, the most important sources are MEDDEV documents and Consensus statements, whereas MDR related guidelines are adopted by the Medical Device Coordination Group (MDCG) with MDCG endorsed documents \cite{eu_commission_web}.  

Finally, regulatory requirements with guidelines and harmonized standards provide a general framework for medical device development while leaving the responsibility to tailor the details of the development process to the manufacturers. 

\subsection{CE marking for medical devices}
A medical device product is considered to have been placed on the market when it is made available for the first time on the Union market \cite{blueguide2016}. Placing on the market is decisive point as there is a legal consequence to comply with all applicable Union legislation at that very moment. The CE marking is the manufacturer's claim that the product meets all relevant regulatory requirements and the process must be completed before the product can be placed on the market. 

To simplify the process, the manufacturer must ensure that \cite{eu_commission_2018}: 
\begin{itemize}
\item The product's intended use fulfills the definition of medical device.
\item The device risk classification has been done correctly based on the intended use.
\item  The correct conformity assessment procedure has been selected.
\item All applicable regulatory requirements have been identified and thoroughly applied.
\item All required technical documentation and the medical device technical file are in place.
\item The QMS has been established (if required by the conformity assessment procedure).
\item The Notified Body has been involved in the assessment (if required by the device risk classification).
\item The new MDR requirements related to unique device identifiers (UDI) have been fulfilled.
\item A declaration of conformity has been signed.
\item CE marking has been applied to the device.
\item Registration to the national competent authority or, in the case of MDR, to EUDAMED database, has been done. 
\end{itemize}

As already mentioned, under MDR almost all MDSW products will require Notified Body involvement in the certification process. Notified Bodies are third party conformity assessment bodies designated by their national competent authority of a member state and they carry out the procedures for conformity assessment within the meaning of applicable Union harmonization legislation \cite{blueguide2016}. 

\section{Mismatch between hardware and software} \label{sec:mismatch}
In this section, we point out a number of regulatory requirement mismatches between physical medical devices and standalone medical device software. 

\subsection{Design Change Approval Process as a barrier to Continuous Deployment}
Continuous maintenance \cite{mikkonen2014}, or the tendency to keep software alive with constant modifications, is nowadays well built into software development practices. DevOps \cite{debois2011}, Continuous Deployment \cite{leppanen2015} and Continuous Software Engineering \cite{fitzgerald2017} are all build on the opportunity to fix bugs, add missing features, and test new, innovative ideas with running software, with real end-users \cite{olsson2014}. However, this approach is clearly in conflict with regulatory framework and Notified Body practices. 

MDD wording related to design change notifications and approvals differs slightly between the different product specific conformity assessment annexes. In the context of software-only medical device, Annex II is used. Annex II, section 4.4 outlines: "\textit{Changes to the approved design must receive further approval from the notified body which issued the EC design-examination certificate wherever the changes could affect conformity with the essential requirements of the Directive or with the conditions prescribed for use of the product. The applicant shall inform the notified body which issued the EC design-examination certificate of any such changes made to the approved design. This additional approval must take the form of a supplement to the EC design-examination certificate}" \cite{mdd}. 

Notified Body Operations Group's (NBOG) Best Practice Guide Guidance for manufacturers and Notified Bodies on reporting of Design Changes and Changes of the Quality System\cite{nbog2014-3} explains the basic principles by which medical device design changes are evaluated by Notified Bodies. In general, NBOG's Best Practice Guides provide guidance on specific aspects related to Notified Body operations and are therefore mainly targeted for Notified Bodies. NBOG BPG 2014-3 differs from this convention as it is clearly targeted also for medical device manufacturers. While NBOG's guidance documents are not legally binding, they are largely adopted by Notified Bodies and organizations responsible for Notified Body designation and control, in order to harmonize and improve the performance of Notified Bodies. As a result, medical device manufacturers are, in practice, expected to follow the principles and requirements outlined in these guidelines in their interaction with their assigned Notified Body. 

NBOG BPG 2014-3 requires manufacturer to have a documented change assessment process with appropriate roles and responsibilities \cite{nbog2014-3}. Furthermore, list of more detailed requirements related to the process is given. For example, the manufacturer shall define change implementation plan in order to monitor the change stages and to ensure compliance to regulatory requirements, update the technical documentation, perform verification and validation activities, classify the change as substantial or non-substantial, and document the decision and timing of the change implementation. 

NBOG BPG 2014-3 also addresses the criteria for deciding if the change is substantial or non-substantial. It is obvious that every change that may affect the conformity with the essential requirements should be treated as substantial. The same applies to changes that affect to product's intended use, risk profile, intended user base, or clinical performance. Changes related to software are specifically addressed in chapter 5.4 and various examples are given. It should be noted that the list of substantial change types is considerably longer than its opposite. To elaborate, the list of non-substantial software changes includes only logic error corrections (that do not pose a safety risk), non-therapeutic and/or non-diagnostic functions (such as printing, faxing, and additional language support), UI appearance changes (with no more than negligible risk of impacting the diagnosis or therapy delivered), and changes to disable a feature that does not interact with other features. It is recommended for manufacturers to contact their Notified Body in any questions related to change classification in order to get mutual understanding. 

According to NBOG BPG 2014-3 \cite{nbog2014-3}, Notified Body must process the change reports with the appropriate actions, for example documenting the assessment related to the product or complete re-assessment of the design documentation of the medical device file, updating the EC certificate if needed, reviewing and updating the contract if needed, and documenting related items that need to be addressed in the next quality management system audit. 

A similar design change approval process can be found from MDR, Annex IX 4.10 \cite{mdr}: "\textit{Changes to the approved device shall require approval from the notified body which issued the EU technical documentation assessment certificate where such changes could affect the safety and performance of the device or the conditions prescribed for use of the device}". As there is currently no MDR specific guidance addressing this issue, it is very likely that the Notified Bodies will continue operating according to their current processes that are based on the guidance given by  NBOG BPG 2014-3. 

It is evident that a process like the one described above is cumbersome for both the manufacturers and Notified Bodies. From the viewpoint of continuous maintenance and agile software development, it can be seen as a huge barrier. Small bug fixes and error corrections can be addressed according to modern software development practices whereas any new safety or clinical performance related features must be approved by regulatory authorities. Therefore, the availability of Notified Body services and resources can be seen as a critical aspect in order to ensure that continuous maintenance activities can be performed in timely manner. Unfortunately, based on authors' experience on the matter, this kind of service level is not often available. 

\subsection{Single-fault condition}
The term single fault condition (SFC) can be found from the MDD \cite{mdd}. The essential requirements in Annex I are grouped under different subheading, and the term appears a total of three times under the following subheadings: 
\begin{enumerate}
\item Construction and environmental properties,
\item Requirements for medical devices connected to or equipped with an energy source, and
\item Protection against electrical risks.
\end{enumerate}
As a result of the wording of the directive, the current interpretation of regulatory authorities is that the SFC is relevant concept exclusively in the context of active physical medical devices. This conclusion is supported by the available guidance documents and IEC 60601 -standard series, where the term is being addressed in the electrical safety context. 

However, this interpretation is not valid in the context of the MDR \cite{mdr} - on the contrary, the term SFC is explicitly connected to software-only device in the Annex I, General safety and performance requirements, in sub-section 17.1: "\textit{Devices that incorporate electronic programmable systems, including software, or software that are devices in themselves, shall be designed to ensure repeatability, reliability and performance in line with their intended use. In the event of a single fault condition, appropriate means shall be adopted to eliminate or reduce as far as possible consequent risks or impairment of performance}". 

From this point of view, there is a clear deviation between the MDD and MDR. This creates the problem of the definition of the term in the software context as the current regulation and guidance is related to physical electrical devices. In software context, there is no generally accepted or generally used definition for the term. Currently, undoubtedly the most important source for the definition is harmonized standard IEC/EN 60601-1 Medical electrical equipment - Part 1: General requirements for basic safety and essential performance, where the term is defined as: "\textit{condition in which a single means for reducing a RISK is defective or a single abnormal condition is present}". The standard also lists few concrete examples related to electrical safety. However, the definition could be used as a first stage when elaborating the exact meaning in the software context.

\subsection{Public Cloud Computing Platforms}
During the last decade, public cloud computing platforms have gained popularity among software development companies. Major  public cloud suppliers like Amazon Web Services (AWS), Microsoft Azure and Google Cloud offer services that consist of hundreds of ready-made software components. While relaying heavily to cloud computing solutions can still be seen as an innovative decision, cloud computing is heading towards mainstream adoption \cite{gartner2019} with massive deployments in various application domains \cite{paladi2017}. However, there are some significant challenges when adopting public 3\textsuperscript{rd} party cloud computing solutions to medical device software development. 

International Medical Device Regulators Forum's (IMDRF) technical document Software as a Medical Device (SaMD): Clinical Evaluation \cite{imdrf2017} describes the process of how standalone medical software should be clinically evaluated. The technical document is wildly adopted by Notified Bodies and used in medical software conformance assessments. According to IMDRF, SaMD clinical evaluation is a planned and systematic process to collect, generate, analyze and assess continuously the clinical data in order to create clinical evidence to prove the clinical association and performance of the SaMD as intended by the manufacturer. The clinical evaluation process consists of three steps, (i) Valid clinical association, (ii) Analytical validation, and
(iii) Clinical validation. In more detail, the analytical validation step includes technical validation. Technical validation should provide objective evidence that the software was correctly constructed and that it correctly and reliably processes data accurately and reproducible. This evidence is usually generated and collected during the verification and validation phase of the software development process. In practice, these technical validation requirements expose the cloud platform to a more detailed analysis. 

The complexity level of the public cloud computing environments creates challenges for the demonstration of conformity. The manufacturer must be able to provide the documentation as required by the regulatory requirements related to adequate performance, risk management and cybersecurity -- the medical device technical documentation should contain clear description of responsibilities between the manufacturer and the cloud platform provider. Furthermore, when processing sensitive health data, the storage requirements related to data location and retention times must be taken into account. 

The cloud-native application (CNA) design seems to correlate highly with cloud-native architecture and design patterns \cite{kratzke2017}. These design guidelines are given by the platform provider when using a specific cloud computing environment. In practice, this  leads to the heavy use of the platform's ready-made components as a part of the medical device software. In fact, in CNA the self-developed part of the software might be significantly smaller than the functionality provided by the existing cloud components. It is undoubtedly a challenge for the manufacturer to create appropriate technical documentation for these 3\textsuperscript{rd} party components and ensure that they remain unchanged after MDSW has been validated. Furthermore, according to MDR Annex I, 17.2 \cite{mdr}, the MDSW must be developed and manufactured taking into account the principles of the development life cycle -- it is highly unlikely that the manufacturer has much control over cloud computing platform supplier's development process. 

If the manufacturer addresses cloud computing platform and its existing components as software of unknown provenance (SOUP) items according to standard IEC/EN 62304 \cite{iec62304}, the specific SOUP requirements of the standard must be fully applied and adequate evidence documented. For example, the manufacturer must specify functional and performance requirements of SOUP items (cl. 5.3.3), verify that the MDSW architecture supports operation of SOUP items (cl. 5.3.6), create the process to evaluate and implement upgrades and fixes of SOUP items (cl. 6.1), consider potential hazardous situations arising from SOUP item failures (cl. 7.1.2), evaluate published SOUP anomaly lists (cl. 7.1.3), analyze the impact of changes to SOUP with respect to safety (cl. 7.4.1) and on existing risk control measures (cl. 7.4.2), and identify and document all SOUP configuration items (cl. 8.1.2). It is evident that this level of scrutiny in documentation requirements is a big challenge when considering 3\textsuperscript{rd} party cloud computing components. 

\subsection{Artificial Intelligence and Machine Learning under the regulatory framework}
With Artificial Intelligence (AI) and Machine Learning (ML) solutions, it is possible to design systems that utilize existing health care data to create new insight for the benefit of the patients. Furthermore, with the recent advances in AI and ML, systems that improve as they perform their critical operations are becoming a reality. For instance, systems that interpret images, recognize speech, and so on can learn on the fly as more and more input data is made available. In contrast, hardware that would improve itself on the fly is still in the laboratories at best, and not ready for prime use. In regulatory actions, there are little possibilities for such silent improvement. In contrast, the regulation aims at keeping the system as it is, with human-validated features \cite{mdr, vogel2011}.

The main difference between the AI/ML system and traditional software product lies, in particular, in the former's ability to adapt and optimize operations in real-time. However, it is possible to design ML application in a way that the system is being trained during the development phase, and the ability to improve is disabled in real-world use. Based on the authors' experience on the matter, this approach is currently indeed promoted amongst EU regulators. As discussed earlier in this Section, all design changes that affect product's clinical performance should be treated as substantial and therefore obtain formal approval according to design change approval process. Clearly, this approach to lock systems self-learning severely limits the efficiency of the ML technology.

While MDR seeks to remedy MDD shortcomings and, as a result, explicitly addresses certain new concepts regarding scientific innovations and emerging manufacturing technologies, AI and ML are not among them. Evidently this is a severe shortcoming in the revised legislation. Against this background, it is surprising to see the theme "Artificial Intelligence under MDR/IVDR framework" addressed in MDCG's ongoing guidance development list under Section 7, New Technologies \cite{mdcg_ongoing_2019}. Guidance on this aspect is highly anticipated by industry stakeholders as there are some significant concerns related to the interpretation of the issue under the MDR. For example, it might be possible to interpret the classification Rule 22 from Annex VIII, Chapter III\cite{mdr} as applicable to AI/ML-based systems, and, as a result, the risk classification of these systems would raise to Class III.  Obviously, such an important principle should be clear for the manufacturer right from the start of the product innovation process in order to ensure lean market entry for AI/ML-based system. 

\subsection{Quality Management System Implementation}
ISO 13485:2016 Medical devices - Quality management systems - Requirements for regulatory purposes \cite{iso13485} is a widely acknowledged standard that defines requirements for the Quality Management System (QMS) for the design and manufacturing processes of medical devices. As it is a harmonized standard, it provides a presumption of conformity with the Essential Requirements \cite{blueguide2016}, which are, in this context,  the requirements outlined in MDD Annex II. 

The QMS processes defined in ISO 13485:2016 are well suited for the organizations performing hardware product manufacturing. However, this is not the case when developing software products where the manufacturing process is more abstract by nature. For example, ISO 13485:2016 requires separate process for transferring the design and development outputs to manufacturing in chapter 7.3.8. In the context of software development, this requirement can be relatively artificial\cite{imdrf2015} as the design and development output is finished software artifact which does not need to be further processed or manufactured. 

Furthermore, ISO 13485:2016 states that infrastructure and work environment must be designed to support the design, development, manufacturing and final inspection of the medical device product according to the product and regulatory requirements. When considering physical manufacturing environment, it is not difficult to understand how lack of infrastructure and work environment control can have an adverse effect to the product quality and, furthermore, to the safety of the product. In contrast, similar physical requirements are generally not applicable in software development context where ordinary office environment is suitable for design and development of the software-only device. 

In contrast to the 2003 revision of the standard, the scope of the 2016 revision has been expanded to be more suitable for any type of organization participating in any activity of the medical device life-cycle. For example, non-applicability option has been extended as not all organizations are expected to perform the full scope of activities addressed in the standard. ISO 13485:2016 states: "\textit{If any requirement in Clauses 6, 7 or 8 of this International Standard is not applicable due to the activities undertaken by the organization or the nature of the medical device for which the quality management system is applied, the organization does not need to include such a requirement in its quality management system. For any clause that is determined to be not applicable, the organization records the justification as described in 4.2.2.}" \cite{iso13485}. However, we could not identify any documented guidance to harmonize regulators' practices related to  non-applicability option assessments as \cite{imdrf2015} does not directly address this matter. 

\section{Discussion} \label{sec:discussion}
Reconsidering regulatory approaches in MDSW development is needed  to ensure compliance with new EU medical device regulations. Furthermore, special attention is required in order to use new software development approaches with medical devices in full \cite{laukkarinen2017,laukkarinen2018}. However, even with the present setup, it is possible to use advanced toolchains, assuming that they have taken regulatory aspects into account \cite{stirbu2018}. Nonetheless, the technical ability to use continuous deployment practices, while simultaneously creating regulatory evidence, can  be  interpreted  to be only partial optimization. As almost all MDSW products will require the Notified Body involvement in the certification process under the new regulations, the co-operational processes between  the medical device manufacturer and the Notified Body play a very central role. Compared to current design change approval procedure, a more streamlined practice is undoubtedly needed to allow continuous software releases. 

MDR brings the term single fault condition explicitly to the software context. Considering the SFC, the definition from the IEC/EN 60601-1, as discussed in Section \ref{sec:mismatch}, could be used to highlight the underlying concept of single fault safe: the product is considered to be safe even if one safety-related requirement or feature fails. In other words, the device's overall risk level needs to remain acceptable even if one feature that involves patient risk, or one risk mitigation fails. In the software and IEC/EN 62304 context, risk control measures for hardware failures and potential software defects must be included to software requirements \cite{iec62304}. As a result, all software requirements that are in some way connected to unacceptable risk - either involving or mitigating one - are affected by SFC requirement.  With this approach, useful and feasible meaning for the term SFC in the software context could be defined, yet the definition would be useful only if adopted by regulatory authorities. 

A critical aspect of using a public cloud computing platform as a technology selection for MDSW is to ensure compliance with regulatory requirements. Major cloud operators are motivated to serve a wide range of different domains, and their platforms are already used by numerous mission-critical sectors, e.g., financial operators. The cloud operators are willing to provide compliant documentation about their services and components, and security and data protection issues are commonly addressed by their infrastructure architecture implicitly. The shared responsibility model of the cloud operator must be well-understood by the manufacturer and appropriately addressed both in product's technical documentation and in the quality management system. For the QMS, the manufacturer should pay a particular focus to cloud governance, configuration management, access control, data encryption, logging and monitoring, incident response, and disaster recovery. The manufacturer must validate the MDSW product built to the cloud environment as usual and the platform operator ensures that their version controlled components and virtual images conform to pre-defined interface specifications and service level agreements. 

When considering the complexity of the MDSW product's regulatory compliance in the cloud environment, it is evident that cross-functional skill-set is needed to provide requisite regulatory evidence. Skillfully designed cloud application architecture could provide specific additional benefits compared to traditional on-premise environments, such as better operational resilience with isolated infrastructure, real-time verification and monitoring, and audit logging. Bearing the mentioned complexities and potential benefits in mind, smaller manufacturers could benefit from the cooperation with each other or with certain suppliers with the ability to deliver a high-quality cloud solutions.

It is somewhat surprising that MDR is not explicitly addressing AI/ML-driven systems. While an increasing amount of new medical device innovations are based on AI/ML systems, the industry stakeholders remain uncertain on how these products will be assessed by the regulatory authorities. Concerning to this issue, U.S. Food and Drug Administration's (FDA) processes are not much more advanced \cite{fda2019} as currently cleared or approved systems typically include only "locked" algorithms and algorithm changes would likely require FDA premarket review. However, on April 2, 2019, FDA published a discussion paper \cite{fda2019} that describes a potential approach to review AI/ML-driven systems. The proposed approach would allow the product to improve in iterative manner while, at the same time, ensure the safety of the patient by enabling real-world performance monitoring. The approach would require manufacturer's commitment for transparency as FDA would expect to get periodic updated on what product changes were implemented. 

In practice, the manufacturers of MDSW must implement a quality management system in compliance with ISO 13485 based on the applicable conformity assessment procedures. In ISO 13485, there are certain elements and procedure requirements that are not well-suited for the manufacturers of software-only devices. In the lack of authoritative guidance on the use of the non-applicability option, the practices may vary amongst assessment bodies and such a variation would jeopardize the equal treatment of manufacturers.

\section{Conclusions} \label{sec:conclusions}
In this paper, we have discussed a few essential EU regulatory requirement mismatches between a physical medical device and standalone software product with intended medical use. The current EU regulatory framework was examined for requirements that do not align well in a software-only context. Special consideration was given to conformity assessment process involving the Notified Body and their assessment practices for the reason of their importance under the new regulations. 

First, we reviewed the change assessment process related to those MDSW products, whose certification process must involve Notified Body assessment under the current  directives. We would highly welcome a more streamlined process under the new regulations as the number of MDSW products affected increases substantially, and the current approach is an obstacle to the efficient use of continuous deployment methods. Second, the term single fault condition was considered in the software context. There is a degree of uncertainty around the term as it remains unexplained by regulatory authorities. However, we proposed a baseline from which the term could be defined. Third, the compliance challenges related to public cloud computing platforms were addressed. The complexity of the cloud environment creates challenges for the manufacturer to collect appropriate regulatory evidence. For this reason, a well-skilled workforce is required. Fourth, the concepts of artificial intelligence and machine learning were reviewed from the viewpoint of regulatory requirements. The identified challenges are related to validation and design change approval processes. MDCG has ongoing guidance development in this field, yet the planned endorsement schedule is open. Last, ISO 13485 standard was researched for it suitability for the manufacturer of software-only devices. Although the standard includes a non-applicability option, the interpretation might vary among the stakeholders. 

Given the current maturity level of the new EU regulatory framework, the future work will include monitoring and analyzing the evolving legislation harmonization work. It is particularly important to gather the new requirements and insight from the MDSW related standards and guidelines yet unpublished. Creating a more streamlined design change approval process would require close co-operation with the regulatory authorities, standardization bodies, and other stakeholders. Although creating such an optimized process may be challenging to do, we believe it would benefit the entire medical device domain. Also, it could be possible to achieve significant clinical benefits if the potential of AI/ML-driven application could be fully utilized in adaptive yet safety ensuring mode. Manufacturers willing to use public cloud computing platforms in their medical device solutions could reduce their regulatory burden and benefit from ready-made, distributed solutions that take into account the regulatory requirements.

\vspace{12pt}

\end{document}